\documentclass[12pt]{article}
\usepackage{graphicx, amsmath, amssymb,color}
\usepackage{hyperref}

\newcommand{\half}{{\frac{1}{2}}}
\newcommand{\mbf}[1]{\mathbf{#1}}

\def\navyblue{\color[rgb]{0,0,0.6}}

\begin{document}

\begin{flushright}
{SLAC--PUB--14154\\
CP3--Origins--2010-22\\
\date{today}}
\end{flushright}

\vspace{40pt}

\centerline{\LARGE  Gauge/Gravity Duality and Hadron Physics}

\vspace{5pt}

\centerline{\LARGE  at the Light-Front}

\vspace{20pt}

\centerline{ 
Guy F. de T\'eramond$^{a}$
and
Stanley J. Brodsky$^{b,c}$
 %
}

\vspace{20pt}

{\centerline{$^{a}${Universidad de Costa Rica, San Jos\'e, Costa Rica}}

\vspace{4pt}

{\centerline{$^{b}${SLAC National Accelerator Laboratory 
Stanford University, Stanford, CA 94309, USA}}

{\centerline{$^{c}$CP$^3$-Origins, University of Southern Denmark, Odense, 5230 M,  Denmark}
}

 \vspace{80pt}

\centerline{\bf Abstract}

\noindent

\vspace{15pt}

We discuss some remarkable features of the light-front holographic mapping of classical gravity in anti-de Sitter space modified by a confining
dilaton background. In particular, we show that  a positive-sign dilaton solution $\exp(+\kappa^2 z^2)$ has better chances to describe the correct 
hadronic phenomenology than the negative solution $\exp{\left(- \kappa^2 z^2\right)}$ extensively studied in the literature. We also show that
the use of twist-scaling dimensions, instead of canonical dimensions, is required to give a good description of  the spectrum and form factors of hadrons. 
Another key element is the explicit connection of AdS  modes of total angular momentum $J$ with the internal structure of hadrons and the proper identification of the orbital angular momentum of the constituents.

\newpage

\section{Introduction \label{int}}

The AdS/CFT correspondence~\cite{Maldacena:1997re} between a gravity or string theory on a higher dimensional Anti-de Sitter (AdS) spacetime 
and conformal gauge field theories  in physical spacetime, modified by color confinement, has led to a semiclassical approximation for strongly-coupled QCD~\cite{deTeramond:2008ht}, which provides
analytical insights into its inherently nonperturbative nature including hadronic spectra, form factors and, very recently, the nonperturbative behavior of the QCD coupling in the infrared region~\cite{Brodsky:2010ur}.

Five dimensional AdS$_5$ spacetime has negative curvature and a four dimensional spacetime boundary.
The most general group of transformations that leave the AdS metric
\begin{equation} \label{AdS}
ds^2 = \frac{R^2}{z^2} \left(\eta_{\mu \nu} dx^\mu dx^\nu - dz^2\right),
\end{equation}
invariant ($R$ the AdS radius), the isometry group, has 15 dimensions, in agreement with the number of generators of the conformal group in four dimensions $SO(4,2)$. This isomorphism is the basic principle
underlying the AdS/CFT approach to conformal gauge theories. Since the metric  (\ref{AdS})
is invariant under a dilatation of all coordinates $x^\mu \to \lambda x^\mu$, $z \to \lambda z$, the variable $z$ acts like a scaling variable in Minkowski space: 
different values of $z$ correspond to different length scales. 

In order to describe a confining theory,
the conformal invariance of AdS$_5$ must be broken. A simple way to impose confinement and  discrete
normalizable modes is to truncate the regime where the string modes can propagate by introducing an IR cutoff at a finite value   
$z_0 \sim 1/\Lambda_{\rm QCD}$. Thus, the ``hard-wall''  at $z_0$ breaks conformal invariance and allows the introduction of the QCD scale  and a spectrum of particle states~\cite{Polchinski:2001tt}.  As first shown by Polchinski
and Strassler,~\cite{Polchinski:2001tt} the AdS/CFT duality, modified
to incorporate a mass scale,  provides a 
nonperturbative derivation of dimensional counting
rules~\cite{Brodsky:1973kr} for the leading 
power-law falloff of hard scattering.

The conformal metric of AdS space can be modified  within the gauge/gravity framework  to simulate  confinement 
forces by  the introduction of an additional warp factor or, equivalently, with a dilaton
background $\varphi(z)$, which introduces an energy scale in the five-dimensional Lagrangian, thus breaking the conformal invariance.
A particularly interesting case is a dilaton profile $\exp{\left(\pm \kappa^2 z^2\right)}$ of either sign, since it 
leads to linear Regge trajectories~\cite{Karch:2006pv} and avoids the ambiguities in the choice of boundary conditions at the infrared wall.  
The modified metric induced by the dilaton can be interpreted in AdS space as a gravitational potential 
for an object of mass $m$  in the fifth dimension:
$V(z) = mc^2 \sqrt{g_{00}} = mc^2 R \, e^{\pm \kappa^2 z^2/2}/z$.
In the case of the negative solution the potential decreases monotonically, and thus an object in AdS will fall to infinitely large 
values of $z$.  For the positive solution, the potential is nonmonotonic and has an absolute minimum at $z_0 = 1/\kappa$.  
Furthermore, for large values of $z$ the gravitational potential increases exponentially, thus confining any object  to distances $\langle z \rangle \sim 1/\kappa$~\cite {deTeramond:2009xk}. 

An important part of this paper is to show that the positive confining dilaton solution $\exp{\left(+ \kappa^2 z^2\right)}$ found in Ref.  \cite{Karch:2006pv}, and subsequently used in \cite{Andreev:2006ct} to describe the confining potential 
between two heavy quarks, has better chances to describe the correct hadronic phenomenology than the negative solution $\exp{\left(- \kappa^2 z^2\right)}$, extensively studied in the literature and known as the ``soft-wall model''.\footnote{The positive dilaton solution was discarded as non-physical in Ref. \cite{Karch:2006pv}
as it leads to a massless $\rho$ meson for the specific AdS/QCD model discussed in Ref.  \cite{Karch:2006pv}.}
Another key element is the explicit connection~\cite{Brodsky:2003px}
 of AdS string modes of total angular momentum $J$ with the internal structure of hadrons and the proper identification of the orbital angular momentum of the constituents~\cite{deTeramond:2008ht}.

Light-front (LF) quantization is the ideal framework for  describing the
structure of hadrons in terms of their quark and gluon degrees of
freedom. Light-front wave functions (LFWFs)  play the same role in
hadron physics that Schr\"odinger wave functions play in atomic physics.
The simple structure of the LF vacuum provides an unambiguous
definition of the partonic content of a hadron in QCD. 
Light-front  holography provides a remarkable
connection between the equations of motion in AdS space and
the Hamiltonian formulation of QCD in physical spacetime quantized
on the light front  at fixed LF time  $\tau = x^+ = x^0 + x^3$~\cite{deTeramond:2008ht}.
 This correspondence provides a direct connection between the hadronic amplitudes $\Phi(z)$  in AdS space  with  LFWFs $\phi(\zeta)$ describing the quark and gluon constituent structure of hadrons in physical spacetime.
The mapping between the LF invariant variable $\zeta$ and the fifth-dimension AdS coordinate $z$ was originally obtained
by matching the expression for electromagnetic (EM) current matrix
elements in AdS space with the corresponding expression for the
current matrix element, using LF  theory in physical spacetime~\cite{Brodsky:2006uqa, Brodsky:2007hb}.   It has also been shown that one
obtains the identical holographic mapping using the matrix elements
of the energy-momentum tensor~\cite{Brodsky:2008pf}, thus  verifying  the  consistency of the holographic
mapping from AdS to physical observables defined on the light front.

\section{Higher Spin Modes in AdS Space}

Our starting point is the Lagrangian for a scalar field in AdS$_{d+1}$ spacetime in presence of a dilaton background field $\varphi(z)$
\begin{equation} \label{S}
S = \int \! d^d x \, dz  \,\sqrt{g} \,e^{\varphi(z)}
  \left( g^{\ell m} \partial_\ell \Phi^* \partial_m \Phi -  \mu^2  \Phi^* \Phi \right)  ,
\end{equation}
where $\varphi(z)$ is a function of the holographic coordinate $z$ which vanishes in the conformal ultraviolet limit $z \to 0$.
The coordinates of AdS are the Minkowski coordinates $x^\ell$ and the holographic variable $z$ labeled $x^\ell = \left(x^\ell \!, z\right)$. Taking the variation of  (\ref{S}) and
factoring out plane waves along the Poincar\'e coordinates,   $\Phi_P(x^\mu,z) = e^{-i P \cdot x} \Phi(z)$,  we obtain the eigenvalue equation
\begin{equation} \label{WES}
\left[-\frac{ z^{d-1}}{e^{\varphi(z)}}   \partial_z \left(\frac{e^\varphi(z)}{z^{d-1}} \partial_z\right) 
+ \left(\frac{\mu R}{z}\right)^2\right] \Phi(z) = \mathcal{M}^2 \Phi(z),
\end{equation}
where 
$P_\mu P^\mu \! = \! \mathcal{M}^2$
is  the invariant  mass  of a physical hadron with four-momentum $P_\mu$.

We define a spin-$J$ mode $\Phi_{\mu_1 \cdots \mu_J}$  with all  indices 
along 3+1 with shifted dimensions
$\Phi_J(z) = ( z/R)^{-J}  \Phi(z)$ and normalization
\begin{equation}  \label{Phinorm}
R^{d - 1 - 2 J} \int_0^{\infty} \! \frac{dz}{z^{d -1 - 2 J}} \, e^{\varphi(z)} \Phi_J^2 (z) = 1.
\end{equation}
The shifted field $\Phi_J$ obeys the wave equation
\begin{equation} \label{WEJ}
\left[-\frac{ z^{d-1 -2 J}}{e^{\varphi(z)}}   \partial_z \left(\frac{e^\varphi(z)}{z^{d-1 - 2 J}} \partial_z\right) 
+ \left(\frac{\mu R}{z}\right)^2\right] \Phi(z) = \mathcal{M}^2 \Phi(z) ,
\end{equation}
which follows from (\ref{WES})
upon mass rescaling $(\mu R)^2 \to (\mu R)^2 - J(d-J)$  and $\mathcal{M}^2 \to \mathcal{M}^2 -  J z^{-1} \partial_z \varphi$.
It is useful to introduce fields with tangent indices 
\begin{equation} \label{eq:ti}
 \tilde \Phi_{i_1 i_2 \cdots i_J}  
 = e_{i_1}^{\ell_1} e_{i_2}^{\ell_2} \cdots e_{i_J}^{\ell_J} \,
 \Phi_{\ell_1 \ell_2 \cdots \ell_J} 
 =  \left(\frac{z}{R}\right)^J  \negthinspace \Phi_{i_1 i_2 \cdots i_J} , 
 \end{equation}
 with scaling  behavior $\tilde \Phi_J(z \to 0) \sim z^\tau$  
 and scaling dimension $\tau$ given by  the relation $(\mu R)^2 = (\tau - J)(\tau -  d + J)$. The vielbein $e^i_\ell$
 is defined by $g_{\ell m} = e^i_\ell e^j_m \eta_{i j}$,
 where  $i, j = 1, \cdots , d+1$ are tangent AdS space indices.

\subsection{Light-Front Holographic Mapping}

In light-front QCD a physical hadron in four-dimensional Minkowski space has four-momentum $P_\mu$ and invariant
hadronic mass states, $P_\mu P^\mu = \mathcal{M}^2$, determined by the 
Lorentz-invariant Hamiltonian equation for the relativistic bound-state system~\cite{Brodsky:1997de}
\begin{equation} \label{LFH}
P_\mu P^\mu \vert  \psi(P) \rangle = \left( P^- P^+ \!  - 
 \mbf{P}_\perp^2\right) \vert \psi(P) \rangle = \mathcal{M}^2 \vert  \psi(P) \rangle.
 \end{equation}
 The hadron  four-momentum generator  is $P = (P^+\!, P^-\!, \mbf{P}_{\! \perp})$, $P^\pm = P^0 \pm P^3$, and
 the hadronic state $\vert\psi\rangle$ is an expansion in multiparticle Fock eigenstates
$\vert n \rangle$ of the free LF  Hamiltonian: 
$\vert \psi \rangle = \sum_n \psi_n \vert n\rangle$.  The internal partonic coordinates of the hadron  are the momentum fractions $x_i = k^+_i/P^+$ and the transverse momenta $\mbf{k}_{\perp i}$, $i = 1, 2,\dots, n$,
where $n$ is the number of partons in a given Fock state. 
Momentum conservation requires 
$\sum_{i=1}^n x_i = 1$ and $\sum_{i=1}^n \mbf{k}_{\perp i}= 0$.  It is useful to employ a mixed 
representation~\cite{Soper:1976jc} in terms of 
 $n-1$ independent momentum fraction variables $x_j$ and position coordinates $\mbf{b}_{\perp j}$, $j = 1, 2, \dots, n-1$,
so that $\sum_{i=1}^n \mbf{b}_{\perp i}= 0$. The relative transverse variables $\mbf{b}_{\perp i}$ are Fourier conjugates
of the momentum variables $\mbf{k}_{\perp i}$. 

The structure of the QCD Hamiltonian equation  (\ref{LFH}) is similar to the structure of the AdS wave equation (\ref{WEJ}); they are both frame-independent and have identical eigenvalues $\mathcal{M}^2$, the mass spectrum of the color-singlet states of QCD, a possible indication of a more profound connection between physical QCD and the physics of hadronic modes in AdS space. However, important differences are also apparent:  Eq. (\ref{LFH}) is a linear quantum-mechanical equation of states in Hilbert space, whereas Eq. (\ref{WEJ}) is a classical gravity equation stemming from general relativity or string theory; its solutions describe spin-$J$ modes propagating in a higher dimensional
warped space. Physical hadrons are composite and thus inexorably endowed of orbital angular momentum. Thus, the identification
of orbital angular momentum is of primary interest in finding a connection between both approaches.  In fact, to a first semiclassical approximation,
light-front QCD  is formally equivalent to the equations of motion on a fixed gravitational background~\cite{deTeramond:2008ht} asymptotic to AdS$_5$ where the prominent properties of confinement are encoded in the dilaton profile $\varphi(z)$. One can indeed systematically reduce  the LF  Hamiltonian eigenvalue Eq.  (\ref{LFH}) to an effective relativistic wave equation~\cite{deTeramond:2008ht}, analogous to the AdS equations, by observing that each $n$-particle Fock state has an essential dependence on the invariant mass of the system $\mathcal{M}_n^2  = \left( \sum_{a=1}^n k_a^\mu\right)^2$ and 
thus, to a first approximation, LF dynamics depend only on $\mathcal{M}_n^2$~\cite{Brodsky:1982nx}.
In  impact space the relevant variable is the boost invariant transverse variable $\zeta$
 which measures the separation of the quark and gluonic constituents within the hadron
at the same LF time and which also allows one to separate the dynamics
of quark and gluon binding from the kinematics of the constituent
internal angular momentum. 
In the case of two constituents, $\zeta = \sqrt{x(1-x)} \vert \mbf{b}_\perp \vert$  where 
$x = k^+/P^+$ is the LF fraction.

Following \cite{deTeramond:2008ht}  we  compute $\mathcal{M}^2$ from the hadronic amplitude
$\langle \psi(P') \vert P_\mu P^\mu\vert\psi(P) \rangle  = \mathcal{M}^2  \langle \psi(P' ) \vert\psi(P) \rangle$,
expanding the initial and final hadronic states in terms of its Fock components.  In the limit of zero quark mass, the longitudinal and transverse
modes decouple and we obtain for a quark-antiquark hadronic bound state the result
\begin{equation} \label{M}  
\mathcal{M}^2   =  \int \! d\zeta \, \phi^*(\zeta) \sqrt{\zeta}
\left( -\frac{d^2}{d\zeta^2} -\frac{1}{\zeta} \frac{d}{d\zeta}
+ \frac{L^2}{\zeta^2}\right)
\frac{\phi(\zeta)}{\sqrt{\zeta}}   \\
+ \int \! d\zeta \, \phi^*(\zeta) U(\zeta) \phi(\zeta) ,
\end{equation}
where all the complexity of the confining interaction terms in the QCD Lagrangian is summed up in the effective potential $U(\zeta)$.
The LF eigenvalue equation $P_\mu P^\mu \vert \phi \rangle  =  \mathcal{M}^2 \vert \phi \rangle$
is thus a light-front  wave equation for $\phi$~ \cite{deTeramond:2008ht}
\begin{equation} \label{LFWE}
\left(-\frac{d^2}{d\zeta^2}
- \frac{1 - 4L^2}{4\zeta^2} + U(\zeta) \right)
\phi(\zeta) = M^2 \phi(\zeta),
\end{equation}
an effective single-variable light-front Schr\"odinger equation which is
relativistic, frame independent and analytically tractable. 

Upon the substitution $z \! \to\! \zeta$  and  
$\phi_J(\zeta)   = \left(\zeta/R\right)^{-3/2 + J} e^{\varphi(z)/2} \, \Phi_J(\zeta)$,
in (\ref{WEJ}), we find for $d=4$ the QCD light-front wave equation (\ref{LFWE}) with effective potential
\begin{equation} \label{U}
U(\zeta) = \half \varphi''(z) +\frac{1}{4} \varphi'(z)^2  + \frac{2J - 3}{z} \varphi'(z) ,
\end{equation}
where the fifth dimensional mass is not a free parameter but scales according to $(\mu R)^2 = - (2-J)^2 + L^2$. 
The LFWFs  are normalized  $\langle \phi \vert \phi \rangle = \int d\zeta \phi^2(\zeta) = 1$.
If $L^2 < 0$ the LF Hamiltonian  is unbounded from below
$\langle \phi \vert H_{LF} \vert \phi \rangle <0$ and thus $\mathcal{M}^2 < 0$;
the particle  ``falls towards the center'' as the effective potential is conformal at small  $\zeta$. 
The critical value corresponds to $L=0$. For $J = 0$ the five dimensional mass $\mu$ 
 is related to the orbital  momentum of the hadronic bound state by
 $(\mu R)^2 = - 4 + L^2$ and thus  $(\mu R)^2 \ge - 4$. The quantum mechanical stability condition $L^2 \ge 0$ is thus equivalent to the
 Breitenlohner-Freedman stability bound in AdS~\cite{Breitenlohner:1982jf}.
The scaling dimensions are $2 + L$ independent of $J$ in agreement with the
twist-scaling dimension of a two-parton bound state in QCD.
It is important to notice that in the light-front the $SO(2)$ Casimir for orbital angular momentum $L^2$
is a kinematical quantity, in contrast with the usual $SO(3)$ Casimir $\ell(\ell+1)$ from non-relativistic physics which is
rotational, but not boost invariant.

\subsection{A Linear Confining Dilaton Background}

A particularly interesting analytical example of a dilaton background 
is that of a Gaussian dilaton profile $\varphi(z) = \pm \kappa z^2$~\cite{Karch:2006pv},
which  corresponds  to a transverse oscillator in the light-front. From (\ref{U}) we obtain for the positive sign solution the effective potential~\cite{deTeramond:2009xk}
$U(\zeta) =   \kappa^4 \zeta^2 + 2 \kappa^2(L + S - 1)$  where $J_z = L_z + S_z$. Equation  (\ref{LFWE}) has eigenfunctions
\begin{equation} \label{phi}
\phi_{n, L}(\zeta) = \kappa^{1+L} \sqrt{\frac{2 n!}{(n\!+\!L\!)!}} \, \zeta^{1/2+L}
e^{- \kappa^2 \zeta^2/2} L^L_n(\kappa^2 \zeta^2) ,
\end{equation}
and eigenvalues
\begin{equation} \label{M2}
\mathcal{M}_{n, L, S}^2 = 4 \kappa^2 \left(n + L + \frac{S}{2} \right).
\end{equation} 
The lowest possible solution for $n = L = S = 0$ has eigenvalue $\mathcal{M}^2 = 0$.
 This is a chiral symmetric bound state of two massless quarks with scaling dimension 2 and size 
 $\langle \zeta^2 \rangle \sim 1/\kappa^2$, which we identify with the lowest state, the pion.
 Thus one can compute the hadron spectrum by simply adding  $4 \kappa^2$ for a unit change in the radial quantum number, $4 \kappa^2$ for a change in one unit in the orbital quantum number and $2 \kappa^2$ for a change of one unit of spin to the ground state value of $\mathcal{M}^2$. Remarkably, the same rule holds for baryons~\cite{deTeramond:2009xk}, thus  predicting the same multiplicity of states for mesons
and baryons, which is observed experimentally~\cite{Klempt:2007cp}.
The LFWFs (\ref{phi}) for different orbital and radial excitations are depicted in Fig. \ref{LFWFs}. Constituent quark and antiquark separate from each other as the orbital and
radial quantum numbers increase.

\begin{figure}[!]
\centering
\includegraphics[angle=0,width=6.8cm]{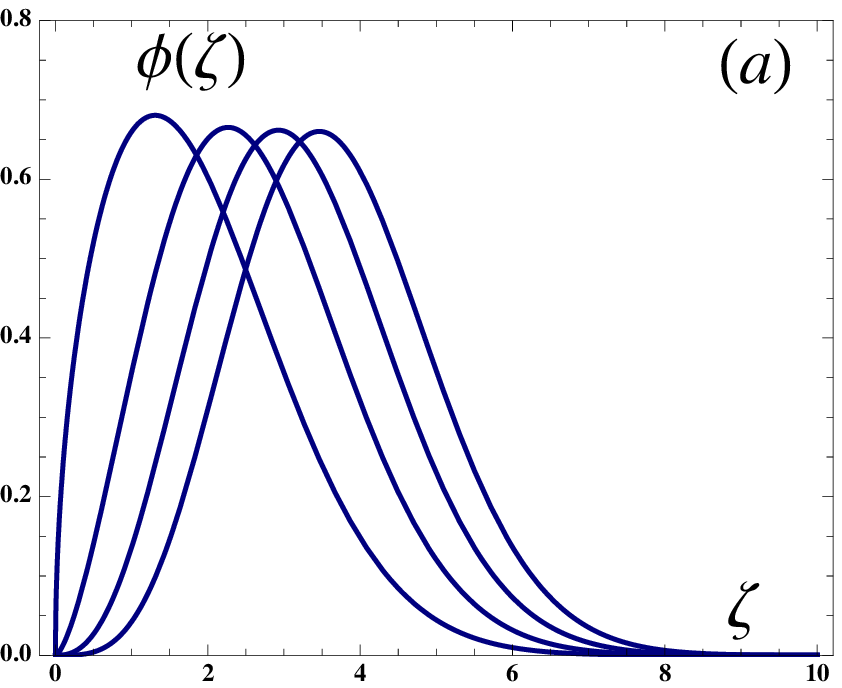} \hspace{20pt}
\includegraphics[angle=0,width=7.0cm]{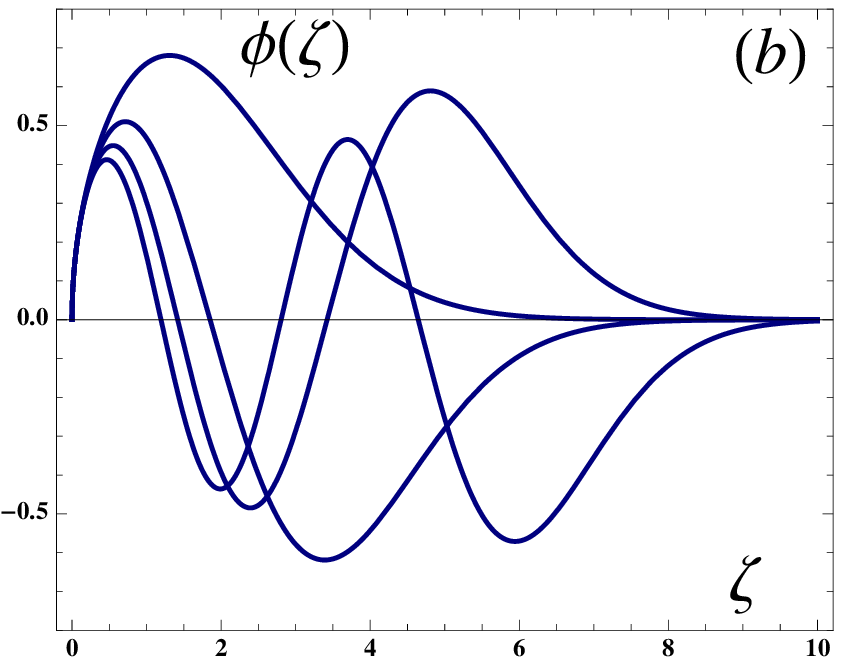}
\caption{Light-front wavefunctions $\phi_{n,L}(\zeta)$  is physical spacetime corresponding to a dilaton $\exp(\kappa^2 z^2)$: a) orbital modes ($n=0$) and b)
radial modes ($L=0$).}
\label{LFWFs}
\end{figure}

Individual hadron states can be identified by their interpolating operator at $z\to 0.$  For example, 
the vector-meson (VM) operator
$\mathcal{O}_{2+L}^\mu = \bar q \gamma^\mu D_{\{\ell_1} \cdots D_{\ell_m\}} q$ with total internal  orbital
momentum $L = \sum_{i=1}^m \ell_i$, is a twist $\tau = 2 + L$ operator with canonical dimension $\Delta = 3 + L$. 
The scaling of  $\tilde \Phi(z)_\mu \sim z^\tau$ at $z \to 0$  is precisely the scaling required to match the scaling dimension of the local vector-meson interpolating operators.   
The spectral predictions for  light VM  states are compared with experimental data 
in Fig. \ref{VMS} for the positive sign dilaton model discussed here.  Only confirmed PDG states~\cite{Amsler:2008xx} are shown.  
 
\begin{figure}[h]
\centering
\includegraphics[angle=0,width=8.6cm]{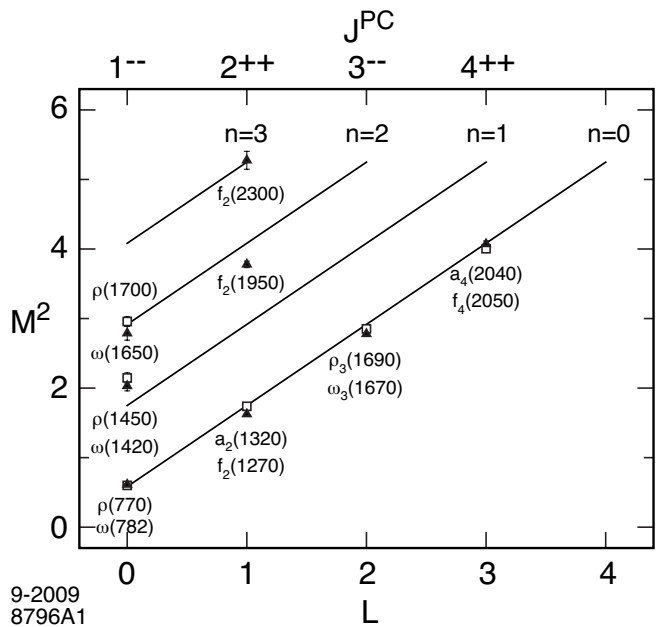}
\caption{Regge trajectories for the  $I\!=\!1$ $\rho$-meson
 and the $I\!=\!0$  $\omega$-meson families for $\kappa= 0.54$ GeV.}
\label{VMS}
\end{figure}

\subsection{Twist versus Canonical Conformal Dimensions}

The short-distance behavior of a hadronic state is characterized by its twist (dimension minus spin) $\tau = \Delta - \sigma$, where $\sigma$ is the sum over the constituent's spin 
$\sigma = \sum_{i=1}^n \sigma_i$. The twist of the interpolating operator ensures dimensional counting rules for form factors and other hard exclusive 
processes~\cite{Brodsky:1973kr}, consistent with conformal invariance at short distances as well as the scaling expected from 
supersymmetry~\cite{Craig:2009rk},
since the scalar,  
the spinor and the gluon fields $G$ all have twist one.  Thus, twist is also equal to the number of partons $\tau = N$.

Consider a hadronic state composed of an arbitrary number of quark and gluons
$\vert q q q \cdots \bar q q \cdots G G \dots \rangle$. As  each individual quark or gluon state has dimension $[\, \vert p_i \rangle ] = [L]$, the hadronic state of momentum $P$, measured at an energy scale $Q$,   should  scale as
$\vert \psi(P)\rangle_Q \sim  \left(1/Q\right)^N$, since  the dimension is set by the energy scale $1/Q$  at large $Q$. The gauge/gravity correspondence implies a duality between
the hadronic state $\vert \psi(P) \rangle$ and the normalizable mode $\Phi_P(x^\mu, z)$ in AdS space. 
Consequently, the relevant scaling dimension of hadronic modes in AdS is dictated by the twist and not the naive conformal  dimension~\cite{Brodsky:2003px}; 
thus $\tilde \Phi(z) \sim z^N$, for twist
$\tau = N$.\footnote{The scaling dimension of a hadronic state with $N$ partons and orbital angular momentum $L$  is $\tau = N + L$.} This is in fact consistent with the scaling of observables.  Consider for example a form factor in AdS 
space~\cite{Polchinski:2002jw, Hong:2004sa}
\begin{equation} \label{F}
F(Q^2) =  R^{3-2J} \int \frac{dz}{z^{3-2J}} \, e^{\varphi(z)} V(Q,z) \, \Phi_J^2(z)  \to \left(\frac{1}{Q^2}\right)^{\tau - 1},
\end{equation}
and the ultraviolet pointlike behavior~\cite{Polchinski:2001ju} responsible for the power law scaling~\cite{Brodsky:1973kr}  is recovered.
The scaling of the form factor at large  
$Q$ follows from integration in the region near $z \sim 1/Q$ where $\tilde \Phi_J(z) = (z/R)^J \Phi(z) \sim z^\tau$. At large $Q$, the bulk-to-boundary propagator $V(Q,z) \sim z Q K_1(z Q)$, and consequently the power-law falloff of the form factors  only depends on the twist-scaling behavior of the hadronic modes and not on  the electromagnetic  current.

Conserved currents are not renormalized and correspond to five dimensional massless fields propagating in AdS according to the relation 
$(\mu R)^2 = (\Delta - p) (\Delta + p -  4)$  for a $p$ form in $d=4$. Thus for an electromagnetic  current the wave equation
\begin{equation} \label{V}
\left[ \frac{z}{e^{\varphi(z)}}\partial_z \left(\frac{e^\varphi(z)}{z} \partial_z\right) - Q^2 \right] A_\mu(Q,z)=0,
\end{equation}
corresponds to $\Delta = 3$ or 1, which are precisely the canonical dimensions of
an EM current and field strength respectively. How 
can we reconcile this assignment 
with the twist-scaling behavior of fundamental hadronic constituents which is required to account for hard scattering?~\cite{Polchinski:2001tt}  For massless quarks, currents do not flip spin. Thus a $\bar q q$ state  produced by an electromagnetic current is an $L_z = \pm 1$ state with the $q$ and $\bar q$ with opposite spins. Consequently, the electromagnetic current is dual to hadronic states with components $L_z=1$ and twist $\tau = 3$.
From  the LF mapping relation $(\mu R)^2 = - (2 - J)^2 + L^2$ there follows that $\mu R = 0$ for $J = L =1$ and  the wave equation (\ref{V}) can also be derived from (\ref{WEJ})
for $J=1$ and $\mathcal{M}^2 \to - Q^2$. The result 
is consistent with conformal dimension $\Delta = 3$, the usual assignment in AdS/QCD models~\cite{Erlich:2005qh, DaRold:2005zs}.

The analysis is similar for a graviton.  The canonical conformal dimension in this case is the dimension 
of the energy-momentum tensor, thus 
$\Delta = 4$. We can identify a $J=2$ field $\Phi_{\mu \nu}$ with an external graviton $h_{\mu \nu}$ propagating in AdS, provided that we take into account the proper normalization of the action for pure gravity. This means that  $h_{\mu \nu} \sim  z^2 \Phi_{\mu \nu}$. From (\ref{WEJ}) we obtain the wave equation
\begin{equation} \label{h}
\left[ \frac{z^3}{e^{\varphi(z)}}\partial_z \left(\frac{e^\varphi(z)}{z^3} \partial_z\right) - Q^2 \right] h_\mu^{\, \nu}(Q,z)=0,
\end{equation}
which coincides with the result obtained by Abidin and Carlson from the linearized gravity action~ \cite{Abidin:2009hr}. This is again consistent with a 
$\bar q q$ state  with $L_z = \pm 2$ and opposite spins produced by a gravitational current.

\subsection{Positive vs Negative Sign Dilaton Background}

\subsubsection{Hadronic Spectrum}

The soft-wall model of Ref. \cite{Karch:2006pv}  uses the AdS/QCD framework of Refs. \cite{Erlich:2005qh} and \cite{DaRold:2005zs}, 
where bulk fields are introduced to match the $SU(2)_L \times SU(2)_R$ chiral symmetry of QCD and its spontaneous breaking, but without an explicit connection to the internal constituent structure of hadrons as done in this article. Instead, axial and vector currents become the primary entities as in an effective chiral theory.  Comparison of both 
approaches is not straightforward and could be misleading, but one would expect that the results are rather similar for both approaches. Comparison of the
results for the $\rho$ principal Regge trajectory of  radial excitations are however significantly different as shown in Fig (\ref{VM}),
 where we compare the predictions of  Ref.  \cite{Karch:2006pv}
with the results from Eq. (\ref{M2}).  This particular example does not
require a discussion of the orbital angular momentum and it is particularly relevant for the computation of hadronic form factors. An AdS mode with a node in the coordinate $z$
should correspond to a radial resonance with a node in the interquark separation. The lowest state, the $\rho(770)$, has no nodes in the wavefunction and corresponds to $n=0$.
For both models we fix the scale $\kappa$ at the $\rho(770)$ mass.

\begin{figure}[h]
\centering
\includegraphics[angle=0,width=6.6cm]{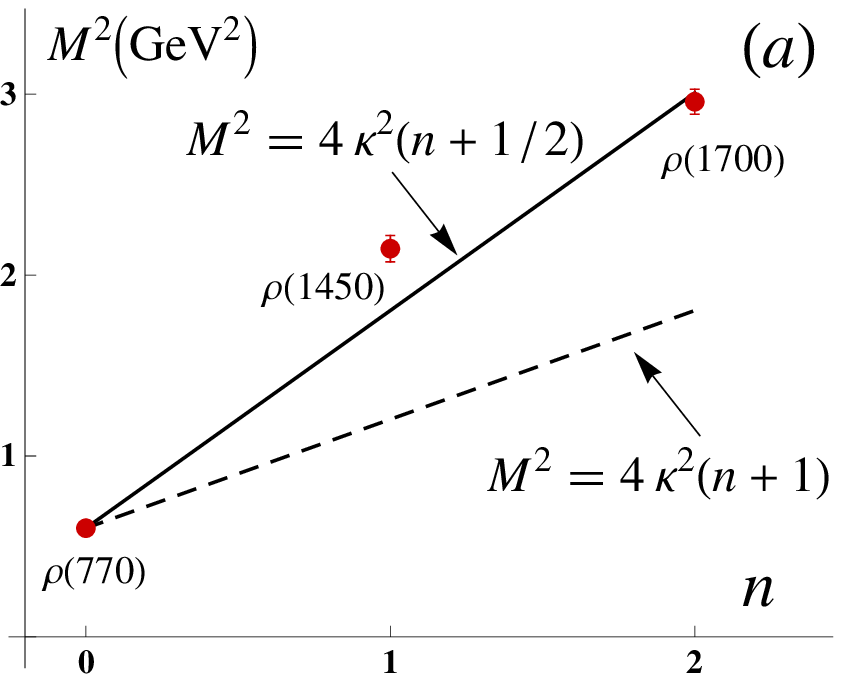} \hspace{30pt}
\includegraphics[angle=0,width=6.8cm]{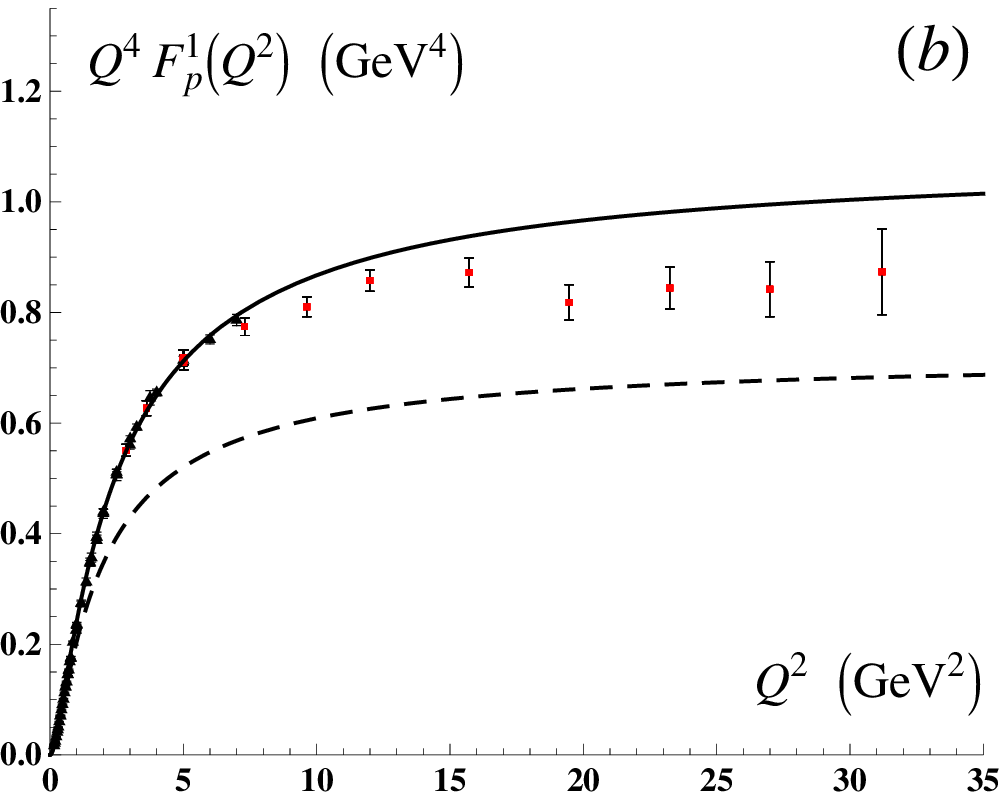}
\caption{Comparison of predictions for hadronic observables in a negative  (dashed line) and  positive  (continuous line) dilaton backgrounds:
a) vector meson radial trajectories and b) space-like scaling behavior of $Q^4 F_1^p(Q^2)$ as function of $Q^2$.
The proton form factor data compilation is from Diehl~\cite{Diehl:2005wq}.
The values of $\kappa$ are  $\mathcal{M}_\rho/2$ ($\varphi = -\kappa^2 z^2$) and $\mathcal{M}_\rho/\sqrt{2}$  ($\varphi=  \kappa^2 z^2)$.}
\label{VM}
\end{figure}

\subsubsection{Hadronic Form Factors}

For the soft wall model of Ref. \cite{Karch:2006pv} the form factor of a hadron of arbitrary twist $\tau$  can be expressed in terms of gamma functions~\cite{Brodsky:2007hb}
by using the integral representation of the bulk-to-boundary propagator found in Ref. \cite{Grigoryan:2007my}.  In absence of anomalous dimensions the twist is an integer,
$\tau = N$ and the result for either sign dilaton
is expressed as an $N-1$ product of poles along the vector radial trajectory~\cite{Brodsky:2007hb}
\begin{equation} \label{FF}
 F(Q^2) =  \frac{1}{\Big(1 + \frac{Q^2}{\mathcal{M}^2_\rho} \Big) 
 \Big(1 + \frac{Q^2}{\mathcal{M}^2_{\rho'}}  \Big)  \cdots 
       \Big(1  + \frac{Q^2}{\mathcal{M}^2_{\rho^{N-2}}}  \Big)} .
\end{equation}  
As expected, the results for the form factor  are sensitive to the location of the VM mass poles  along the Regge trajectory. This is more noticeable in the time-like region which is particularly sensitive to the detailed pole structure and the interference effects from the phase of the amplitude.
We show in Fig.  \ref{VM} the results for the 
Dirac spin-non-flip proton form factor.  Since the proton is twist three the results follow from the first two terms in Eq. (\ref{FF}).
The dashed line corresponds to the minus sign dilaton $\varphi = - \kappa^2 z^2$ with $M^2 = \kappa^2(n+1)$ and the continuous line to the positive sign dilaton
$\varphi = \kappa^2 z^2$ with VM masses $M^2 = 4\kappa^2( n + 1/2)$. For both models the scale $\kappa$ is fixed  at the $\rho(770)$ mass. The positive dilaton solution
gives a better result when we use the mass eigenvalues of the
Hamiltonian equation which are close to the observed masses, thus shifting the poles from twist-three (the dimension required to have zero fifth dimensional mass in the conserved current) to twist two, the twist of a two-component object. How can this procedure be justified?

In the limit of zero quark mass, states such as $e^+  e^-$, $\bar q q$ or $G G$  are produced with opposite spin and non-zero orbital angular momentum.  The bound states, however, can have a zero $L_z$ component, which mix
with states with non-zero orbital.  For example, the 
vector meson $\rho(770)$ state has multiple components $S = 1, L_z = 0$ as well as  $S = 0, L_z = \pm 1$.  The mass of the state can be read off from the lowest twist ($L_z=0$) bound state.  For example, the solution  for a spin-$\half$ field in AdS has two components: a plus component  $\psi_+$ which represent a state with $S_z = \pm \half, L_z = 0$, and a  minus component  $\psi_-$ which represents a state with $S_z = \mp \half, L_z = \pm 1$. Since both components mix in the AdS wave equation they have the same mass: the value of the mass eigenstate is the same for the lower twist $L_z=0$ component than for the higher $L_z = \pm 1$ component~\cite{deTeramond:2009xk}. Thus to take into account the different components of  a VM state, one requires a multiple-component 
equation in AdS space, as for the spin-$\half$ case. This multiple-component equation (a Kummer-Duffin-Petiau-like equation)  for vector mesons has not been derived, to our knowledge, in AdS space, 
since one usually compute the mass spectrum from a single-component wave equation (like Eq. (\ref{WEJ})), where the eigenmass corresponds to the lowest ($L_z = 0$) state. On the other hand, the mass poles in the Green's function 
for the current propagator of a VM AdS field $A(z)$ is~\footnote{In terms of the Green's function (\ref{GF}) the  bulk-to-boundary propagator $V(q,z)$ in (\ref{F})  is
$V(q,z) =  V(q,0) \lim_{z' \to 0}    e^{\varphi(z')} \frac{R}{z'} \partial_{z'}   G(z, z'; q)$~\cite{Hong:2004sa}.}
\begin{equation} \label{GF}
G(z,z';q) = R \sum_n \frac{e^{\kappa^2 z'^2}}{z'}  \,
\frac{A_n(z') A_n(z)}
{M_n^2 - q^2- i \epsilon},
\end{equation}
corresponding to the higher-twist component and not to the physical mass (which corresponds to the lowest-twist component in the eigenvalue equation) when proper component mixing is allowed in the equations of motion. Consequently the location of the poles has to be shifted to their lowest-twist (physical) mass to describe
correctly the form factors in a positive dilaton background.

\subsubsection{Nonperturbative QCD coupling}

Very recently we have examined with Alexandre Deur the behavior of  nonperturbative  effective couplings in QCD from the perspective of light-front 
holography~\cite{Brodsky:2010ur}. The infrared (IR) results for the strong coupling are markedly different according to the sign of the dilaton chosen. The positive
dilaton $\varphi(z) = \kappa^2 z^2$ leads to an IR fixed-point. In contrast, the negative solution $\varphi(z) = - \kappa^2 z^2$ leads to a  coupling which blows up in
the  IR. Following~\cite{Brodsky:2010ur}, we consider a five-dimensional gauge field $F$ propagating in AdS$_5$  in the presence of  dilaton $\varphi(z)$.
At quadratic order in the field strength the action is 
\begin{equation}
S =  - {1\over 4}\int \! d^4x \,  dz  \sqrt{g} \, e^{\varphi(z)}  {1\over g^2_5} \, F^2,
\label{eq:action}
\end{equation}
where we identify the prefactor  $g^{-2}_5(z) =  e^{\varphi(z)}  g^{-2}_5$,
as the effective coupling of the theory at the length scale $z$. 
The coupling $g_5(z)$  incorporates the nonconformal dynamics of confinement. The five-dimensional coupling $g_5(z)$
is mapped,  modulo a  constant, onto
the Yang-Mills (YM) coupling $g_{\rm YM}$ of the confining theory in physical space-time using light-front holography. One  identifies $z$ with the invariant impact separation variable $\zeta$: $g_5(z) \to g_{\rm YM}(\zeta)$. Thus 
$\alpha_s^{AdS}(\zeta) = g_{\rm YM}^2(\zeta)/4 \pi \propto  e^{-\kappa^2 \zeta^2}$.  The physical coupling measured at the scale $Q$ is the two-dimensional Fourier transform
of the  LF transverse coupling $\alpha_s^{AdS}(\zeta)$. Integration over the azimuthal angle gives
 \begin{equation} \label{eq:2dimFT}
\alpha_s^{AdS}(Q^2) \sim \int^\infty_0 \! \zeta d\zeta \,  J_0(\zeta Q) \, \alpha_s^{AdS}(\zeta)  \sim e^{- Q^2 /4 \kappa^2}.
\end{equation}

\begin{figure}
\includegraphics[angle=0,width=7.4cm]{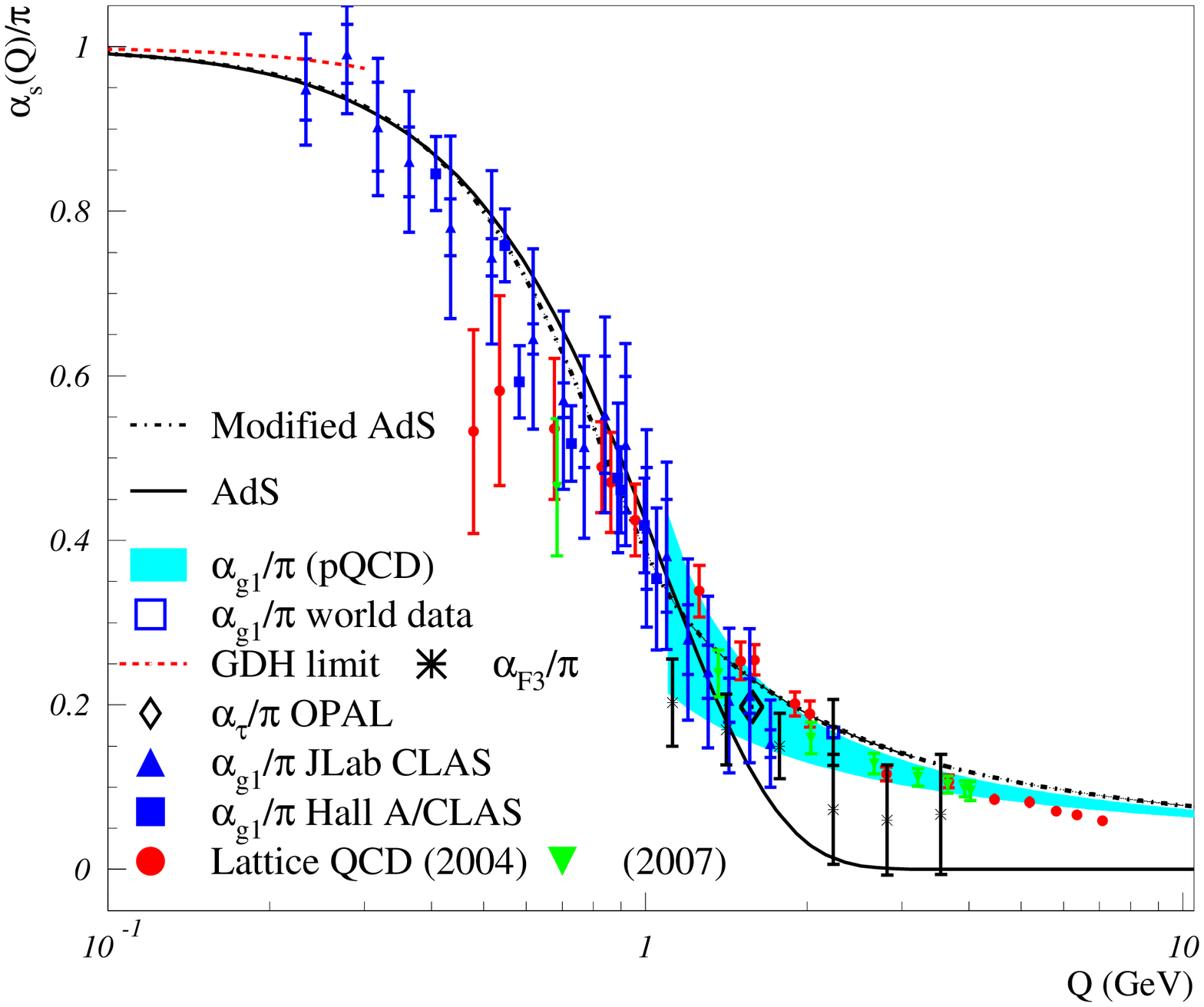} \hspace{10pt}
\includegraphics[angle=0,width=7.4cm]{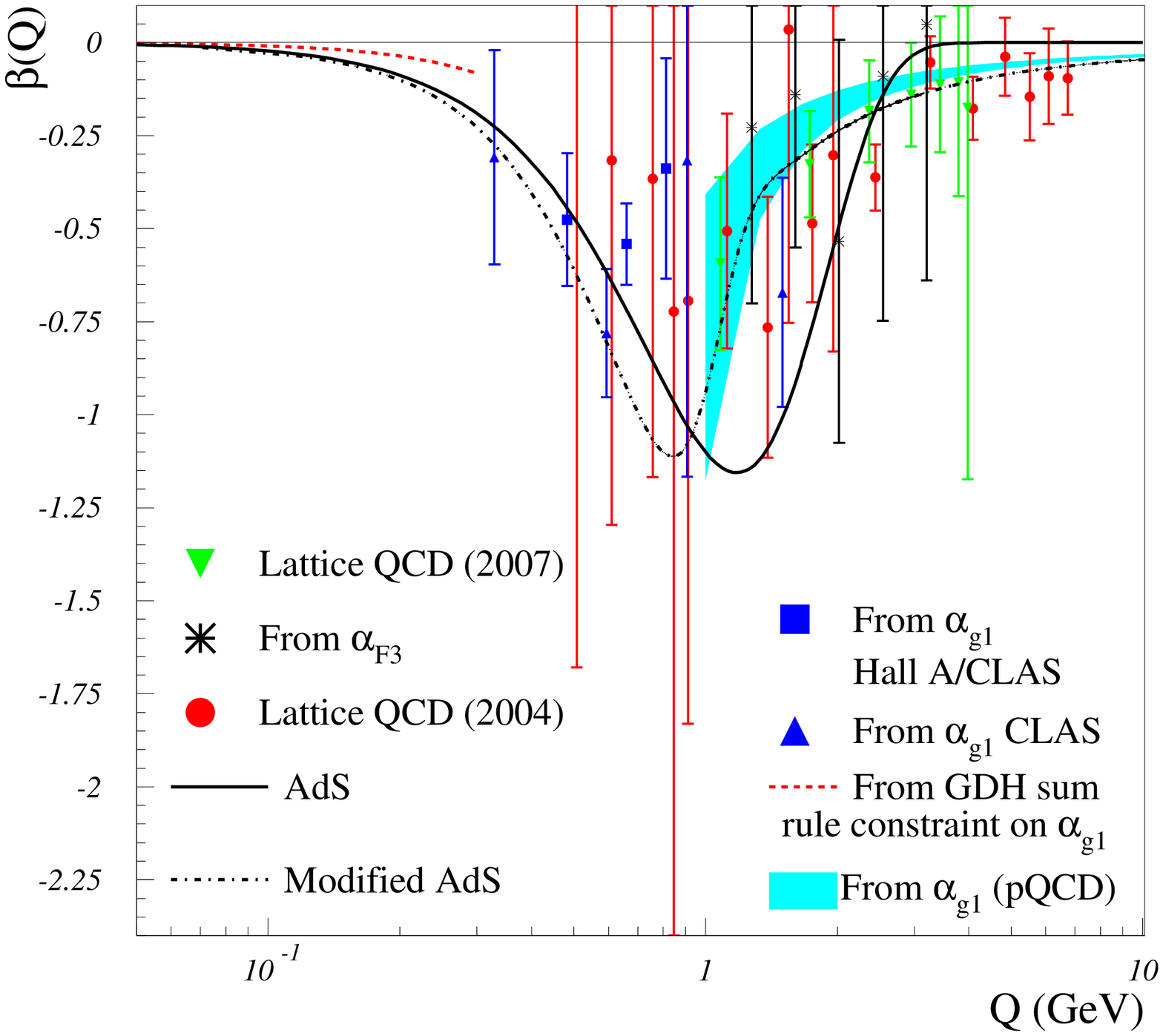} 
\caption{\label{alphas}The strong coupling (left) and $\beta$ function (right) from LF holographic mapping  (continuous line) for  $\kappa = 0.54 ~ {\rm GeV}$ are compared with effective QCD results
extracted from different observables and lattice simulations. Details of the comparison and normalization used are given in Ref. \cite{Brodsky:2010ur}.}
\end{figure}

The strong coupling  $\alpha^{AdS}(Q^2)$ is compared in Fig. \ref{alphas} with  experimental and lattice data.
The falloff  of $\alpha_s^{AdS}(Q^2)$  at large $Q^2$ is exponential: $\alpha_s^{AdS}(Q^2) \sim e^{-Q^2 /  \kappa^2},$ rather than the perturbative QCD (pQCD) logarithmic falloff,
since effects from gluon creation and absorption are not included in the semiclassical theory.
The corresponding beta function in Fig. \ref{alphas} is conformal in the infrared and ultraviolet (UV) regions. It  becomes significantly  negative in the nonperturbative regime $Q^2 \sim \kappa^2$, where it reaches a minimum, signaling the transition region from the IR conformal region, characterized by hadronic degrees of freedom,  to a pQCD conformal UV  regime where the relevant degrees of freedom are the quark and gluon constituents.  The  $\beta$ function is always negative; it vanishes at large $Q^2$ consistent with asymptotic freedom, and it vanishes at small $Q^2$ consistent with an infrared fixed point.

\section{Conclusion}

Light-front  holography provides a direct correspondence between an effective gravity theory defined in a fifth-dimensional warped space and a physical description of hadrons in  $3+1$ spacetime. The relativistic light-front  wave equations which follow from the semiclassical approximation to the gauge/gravity correspondence in light-front QCD provide
remarkably successful predictions for the light-quark meson and baryon spectra as 
a function of hadron spin, quark angular momentum, and radial quantum number. 
The predictions for form factors are also remarkably
successful, and the predicted power law fall-off agrees with dimensional counting rules as required by conformal invariance at small distances. The use of twist-scaling 
dimensions and the proper identification of the orbital angular momentum of the constituents are key elements to describe
the observed hadronic data. As in the Schr\"odinger equation, the semiclassical approximation to light-front QCD described in this paper does not account for particle
creation and absorption; it is thus expected to break down at short distances
where hard gluon exchange and quantum corrections become important. 
However, one can systematically
improve the semiclassical approximation, for example by introducing nonzero quark masses and short-range Coulomb
corrections.


\vspace{30pt}

\noindent {\bf \Large Acknowledgments}

\vspace{12pt}

Invited talk presented by GdT at XI Hadron Physics,  21-26 March  2010, Maresias Beach, S\~ao  Paulo, Brazil. GdT is grateful to the organizers 
for their outstanding hospitality.  We  thank A. Deur,  H. G. Dosch and J. Erlich,
for helpful comments and collaborations.  This research was supported by the Department
of Energy  contract DE--AC02--76SF00515.  SJB also thanks the Hans Christian Andersen Academy and CP$^3$-Origins for their support.


\end{document}